\documentstyle[aps,prl,multicol,epsf]{revtex}
\global\firstfigfalse
\global\firsttabfalse
\renewcommand{\narrowtext}{\begin{multicols}{2} \global\columnwidth20.5pc}
\renewcommand{\widetext}{\end{multicols} \global\columnwidth42.5pc}
\input{epsf.tex}

\def\top#1{\vskip #1\begin{picture}(290,80)(80,500)\thinlines \put(
65,500){\line( 1, 0){255}}\put(320,500){\line( 0, 1){
5}}\end{picture}}
\def\bottom#1{\vskip #1\begin{picture}(290,80)(80,500)\thinlines \put(
330,500){\line( 1, 0){255}}\put(330,500){\line( 0, -1){
5}}\end{picture}}

\def\Tr{\mbox{Tr}}
\def\dif{\stackrel{\leftrightarrow}{\nabla}}

\newcommand{\bleq}{\ifpreprintsty
                   \else
                   \end{multicols}\vspace*{-3.5ex}{\tiny
                   \noindent\begin{tabular}[t]{c|}
                   \parbox{0.493\hsize}{~} \\ \hline \end{tabular}}
                   \fi}
\newcommand{\eleq}{\ifpreprintsty
                   \else
                   {\tiny\hspace*{\fill}\begin{tabular}[t]{|c}\hline
                    \parbox{0.49\hsize}{~} \\
                    \end{tabular}}\vspace*{-2.5ex}\begin{multicols}{2}
                    \fi}
\newcommand{\bcols}{\ifpreprintsty\else\begin{multicols}{2}\fi}
\newcommand{\ecols}{\ifpreprintsty\else\end{multicols}\fi}

\begin{document}
\bibliographystyle{prsty}
\title{Landauer Conductance without Two Chemical Potentials}
  
\draft

\author{Alex Kamenev$^1$, and Walter Kohn$^2$}
\address{$^1$Department of Physics, Technion, Haifa 32000, Israel.\\
$^2$Department of Physics, 
 University of California Santa Barbara, CA 93106-4030.
  \\
  {}~{\rm (\today)}~
  \medskip \\
  \parbox{14cm} 
    {\rm We present a theory of the four--terminal conductance 
for the multi--channel tunneling barrier, which is based on the self-consistent 
solution of Shr\"{o}dinger, Poisson and continuity  equations.   
We derive new results for the case of a
barrier  embedded in a long wire with and without disorder.
We also recover  known expressions for the conductance of the barrier placed 
into a  ballistic  constriction. Our approach 
avoids a problematic use of  two chemical potentials in the same system. 
     }\bigskip \\ }

\maketitle

\narrowtext

\section{Introduction}
\label{s1}

Tunneling conductance through a potential barrier has been  the subject of 
intense study since the original Landauer paper \cite{Landauer57} in 1957 
(see Ref.  \cite{Imry86} for reviews). 
It was recognized that the {\em two--terminal} 
measurements of the tunneling conductance, $g$, 
may be described by the Landauer formula 
\begin{equation} 
g=\frac{e^2}{\pi\hbar}
\sum\limits_{i} |t_{i}|^2\, ,
                                                              \label{q1}
\end{equation}  
where the $t_i$ are the eigenvalues of 
a barrier transmission matrix. 
The term ``two--terminal'' implies that the voltage drop 
is measured between the same two contacts, sufficiently far from the barrier, 
which are used to pass a current. Actually  Landauer's original 
expression \cite{Landauer57}, 
for the single channel case, had the different form
\begin{equation} 
\tilde g=\frac{e^2}{\pi\hbar}\,  \frac{|t|^2}{|r|^2}\, ,
                                                              \label{q2}
\end{equation}  
where $r$ is the reflection coefficient of the barrier, obeying 
$|r|^2 + |t|^2=1$. Equation (\ref{q2}) is intuitively appealing since it leads 
to  zero resistance for the fully transparent ($|t|=1$) barrier. Only after 
the work of Thouless \cite{Thouless81} and Imry \cite{Imry86} 
was it realized that Eq.~(\ref{q2}) 
implicitly relies on the ability to measure the voltage drop right across the 
tunneling barrier. One should thus assume the existence of another pair of 
contacts used solely as   potential (as opposed to current) probes -- 
a {\em four--terminal} measurement in  modern terminology. 
The difference for the single channel 
case, 
$g^{-1} - \tilde g^{-1} = \pi\hbar/e^2$, was interpreted  as an effective  
resistance of the perfect wire, which connects the barrier with the  
current contacts. 

The two--terminal result, Eq.~(\ref{q1}), may be derived from the Kubo 
formula for non--interacting electrons  
\cite{Lee80,Economou81,Stone89}. Its general form is unchanged 
by the Hartree screening of the barrier \cite{Serota}, 
which will, however, alter the transmission 
coefficients. 

Thouless \cite{Thouless81} confirmed Eq.~(\ref{q2}) for an electron gas,
interacting on the Hartree level, by imposing charge neutrality in the wire 
on  long length scales.  In fact, the four--terminal expression, 
Eq.~(\ref{q2}), is the result of a self--consistent calculation of the 
charge density induced near the barrier. Although this charge does not 
alter the divergence-less current, it does determine the voltage drop across 
the barrier region. 
To illustrate this point let us consider a long 
single--channel wire with an embedded barrier. Suppose 
that the wire is placed in a time--dependent, 
spatially uniform external electric field 
$E(t)=E^{ext}\cos\omega t$. This can be done, for example, by  
bending the wire into a ring, which encloses a time--dependent magnetic 
flux. Far from the barrier the electron current corresponds to  
pure acceleration, it is position independent and given by 
$j(t) = e^2 n /(m\omega) E^{ext} \sin\omega t$, 
where $n$ is the electron density. Close to the barrier, however, 
the current is space dependent and charges accumulate 
on 
both sides of the barrier. These induce an additional electric field which 
contributes to the voltage drop, $V$, across the barrier. 
Calculations given below lead to the result 
$V=E^{ext} \sin(\omega t) 2v_F|r|^2/(|t|^2\omega)$. 
Taking into account that in one dimension $n=2 m v_F/(\pi\hbar)$, 
where $v_F$ is the 
Fermi velocity, one recovers Eq.~(\ref{q2}) for the conductance, in agreement 
with Thouless's conclusion.

While the single--channel four--terminal expression,  Eq.~(\ref{q2}), is 
generally agreed upon, the generalization to the multi--channel case has been 
a subject of 
controversy. P.~W.~Anderson {\em et. al.} \cite{Anderson80}  
proposed to add the conductances of channels in parallel. 
Langreth and Abrahams \cite{Langreth81} assumed that 
the chemical potential, say on the left of the barrier, 
is determined by the mean 
chemical potential of all left and right moving electrons. The difference 
between mean left and mean right chemical potential was identified with the 
voltage drop. Calculating the conductance required solving   
a large system
of equations. Their approach assumes the existence of strong 
independent equilibration 
among the left- and right-moving particles, and the absence of any cross--talk 
between  the left and right movers. As was first remarked by 
B\"{u}ttiker, Imry, Landauer and Pinhas (BILP) \cite{Buttiker85}, 
there is no  universal expression 
for the conductance in the multichannel case; it depends on details of the 
electron distribution function in the leads. The latter may depend on 
specific geometry, relaxation mechanisms, etc. \cite{Landauer98}. 
BILP considered a particular 
case of scattering--free leads connected to  two ideal thermal reservoirs. 
Each reservoir is  
taken to be in thermal equilibrium with different chemical potentials 
on the left and  right of the barrier. The distribution functions of 
the left and right    moving carriers are strictly imposed by the reservoirs 
from which they  originated. Thus on each side of the barrier left and right 
moving electrons occupy Fermi hemispheres of different radii. 
The potential drop across the barrier is taken to be the 
difference of effective chemical potentials of the leads, which are 
determined from the postulated 
condition of having equal number of particles above and 
holes below them. For the case of no channel mixing, 
$t_{ij}= t_i \delta_{ij}$ and $r_{ij}= r_i \delta_{ij}$, BILP obtained 
\begin{equation} 
\tilde g_{BILP}=\frac{e^2}{\pi\hbar} \, 
\frac{\sum\limits_i v_i^{-1} \sum\limits_i |t_i|^2}
{\sum\limits_i v_i^{-1} |r_i|^2}\, ,
                                                              \label{q3}
\end{equation}  
where $i$ runs over the open channels with velocity $v_i$. 
The same result was 
obtained earlier by Azbel \cite{Azbel81}, using a slightly different 
argumentation. For a single channel Eq.~(\ref{q3}) reduces to the 
old Landauer result, Eq.~(\ref{q2}), whereas for a very high barrier, 
$|r_i|\approx 1$, there is no difference between the four--terminal result, 
Eq.~(\ref{q3}), and the two--terminal one, Eq.~(\ref{q1}).   

In this paper we develop a logically clean 
formulation of the four--terminal conductance 
which avoids the problematic introduction of different chemical 
potentials in  various parts of the same system. Instead, we consider a pure 
Hamiltonian formulation of the problem with the external bias applied by 
a time dependent vector potential through a macroscopically large loop containing a 
barrier. We solve  self-consistent equations for the electric field 
on the Hartree (or RPA) level to determine   the voltage drop across the barrier.
As a microscopic ingredient we need only an equilibrium linear conductance kernel,
which is derived using the Kubo formula. Our 
formulation of the problem also allows us to tackle regimes not considered 
previously. As an example we consider  a barrier embedded 
into a long multi--channel wire of uniform cross-section (a geometry different from 
that considered by BILP). 
  If the wire length, $L$, exceeds $v_F/\omega$ 
($v_F$ is a Fermi velocity), the  electron 
distribution function is  a shifted (accelerated) 
Fermi sphere, rather than the BILP distribution. 
This results in a four--terminal conductance which is different from the BILP 
expression. 
For example in  the low barrier limit, $|r_i|\ll 1$, we find   
\begin{equation} 
\tilde g=\frac{e^2}{\pi\hbar}\,  
\frac{\left[ \sum\limits_i v_i \right]^2 }
{\sum\limits_i v_i^{2} |r_i|^2}\, ,
                                                              \label{q4}
\end{equation}  
which is quite different from Eq.~(\ref{q3}) in the same limit. For 
one thing, we do not expect  discontinuities of conductance as function of  
the Fermi energy at the values where velocity of one of the channels is zero. 
In the opposite limit of a high barrier, $|t_i|\ll 1$, we obtain Eq.~(\ref{q1}),
in agreement with BILP and other approaches. 
Another regime we are able to treat is a barrier embedded in 
a disordered wire with $L \gg l^{el}$, where  
$l^{el}$ is an elastic mean free path. 
In this case, the momentum equilibration which 
takes place in the leads due to  strong 
impurity scattering. We find that the same result, Eq.~(\ref{q4}), holds 
for the case of disordered leads. 
That does not mean that the BILP expression, Eq.~(\ref{q3}), is incorrect. We have 
recovered it, together with the conditions for its applicability, for a barrier 
embedded in a  ballistic constriction, $L < v_F/\omega,l^{el}$, between 
two wide reservoirs. 
Our procedure is quite different from that of BILP. In particular we do not 
require consideration of an unphysical, space dependent 
chemical potential --- it may be defined using only standard equilibrium 
thermodynamics  along with Poisson's and linear response equation.

The outline of the paper is as follows: in Sec. \ref{s3} we formulate 
self--consistent equations for the induced electric field and charge
density. In 
Sec. \ref{s2} we derive a microscopic 
expression for the {\em non--local} conductance in the case of the clean 
wire. We then solve the self--consistent field equations and obtain 
the four--terminal conductance in a number of regimes. 
Sec. \ref{s5} deals with
a barrier with disordered leads. 
In Sec. \ref{s7} we treat an adiabatic constriction geometry. 
Finally, in 
Sec. \ref{s4} we discuss the results, their domains of validity 
and  possible generalizations.

\section{Self--consistent Field Equations}
\label{s3}

Let us consider a multi--channel metallic wire of uniform 
cross-section, $S$, along the 
$z$--direction. We assume that in a vicinity of $z=0$ there is a 
potential barrier (or more generally some localized scattering region) 
across the wire. A weak uniform applied electric field, $E^{ext}(\omega)$, 
with  frequency $\omega$  creates a current density, 
which in the linear response regime has the form   
\begin{equation} 
j(r,\omega) = \int\! d r' g(r,r',\omega)\cdot E(r',\omega) \, .
                                                              \label{w1}
\end{equation}
Here the integration runs over the volume of the wire and $E(r',\omega)$ is a 
{\em total} electric field at the point $r'$. The non--local 
linear conductivity kernel, $g(r,r',\omega)$,  may be calculated  
using the Kubo formula.

We assume that, due to the quasi--one--dimensional geometry 
of the system, the electric field is practically  
independent of the transverse coordinates, 
$E(r,\omega)\approx E(z,\omega)$ and only the total current  
$I(z,\omega) = \int_S dS j(r,\omega)$ is measurable experimentally.  
The total current, 
$I(z,\omega)$, and the electric field, $E(z,\omega)$, are linearly connected 
through the {\em conductance} kernel, $g(z,z',\omega)$. 
Employing the  Kubo formula \cite{Lee80,Economou81,Stone89} and disregarding 
the  e--e interactions in the kernel $g(z,z',\omega)$ 
(which is equivalent to the time--dependent Hartree approximation), one 
may express the  non--local conductance as \cite{foot1} 
\begin{eqnarray} 
g(z,z',\omega) =&& \frac{e^2}{2im^2\hbar}\!  
\int\!\! d\epsilon  \,
\frac{f(\epsilon_+) - f(\epsilon_-)}{\omega} \times
                    \label{w2} \\
&&\sum\limits_{ij}
G^{R}_{ij}(\epsilon_+;z,z')\!\! \dif_z \dif_{z'} \!\!
G^{A}_{ji}(\epsilon_-;z',z)  \, ,  
                                                              \nonumber
\end{eqnarray}
where $A\dif_zB\equiv A\nabla_z B - \nabla_zA B$,  
$\epsilon_{\pm}=\epsilon\pm\omega/2$, 
$f(\epsilon_{\pm})$ is the Fermi 
function and  $G^{R(A)}(\epsilon)$ is the retarded (advanced) Green's
function of the system at energy $\epsilon$ in the transverse channel basis, 
\begin{equation}
G_{ij}(\epsilon;z,z') = 
\int\limits_{S(z)}\!\!\! dS\int\limits_{S(z')}\!\!\! dS'\, 
\xi_i^*(x,y)G(\epsilon;r,r') \xi_j(x',y')  \, .  
                                                              \label{w5}
\end{equation}
Here $\xi_i(x,y)$ is the eigenfunction of the $i$-th transverse channel with 
energy $\epsilon_i$.
It will 
sometimes be convenient to divide the conductance into  two parts,
\begin{equation} 
g(z,z',\omega) = g_0(z-z',\omega) + g_b(z,z',\omega)\, ,
                                                              \label{n1}
\end{equation}
where $g_0(z-z',\omega)$ is the translationally invariant conductance of 
the wire without the barrier and $g_b(z,z',\omega)$ is the contribution 
associated with the scattering by the barrier.

The linear response relation, Eq.~(\ref{w1}), together with the continuity 
and  Poisson's equations, constitute a closed system of equations for  three 
unknown quantities: the current  $I(z,\omega)$, the electric field 
$E^{ind}(z,\omega)$ and the charge density $\rho(z,\omega)$ all induced by the 
uniform external field, $E^{ext}(\omega)$, 
\begin{eqnarray} 
&&I(z,\omega) = \int\! d z' g(z,z',\omega) 
(E^{ext}(\omega) + E^{ind}(z',\omega))\, ,
                                                            \label{e1a} \\
&&\partial_z I(z,\omega) = i\omega \rho(z,\omega)\, ,
                                                            \label{e1b}\\
&&\partial_z E^{ind}(z,\omega) = 4\pi\rho(z,\omega)/S\,\, .
                                                             \label{e1c}
\end{eqnarray}
Eliminating $I(z,\omega)$ and $\rho(z,\omega)$, 
one obtains the following integral equation for $E^{ind}(z,\omega)$ 
\begin{eqnarray} 
&&\frac{S}{4\pi}\, \partial_z E^{ind}(z,\omega) -
\frac{1}{i\omega} \int\!\! d z' \partial_z g(z,z',\omega) E^{ind}(z',\omega) 
           \nonumber  \\
&&= \rho^{(0)}(z,\omega) \, ,     
                                                              \label{e2}
\end{eqnarray}
where $\rho^{(0)}(z,\omega)$ is the bare (unscreened) charge density produced by 
the external electric field near the barrier 
\begin{equation} 
\rho^{(0)}(z,\omega) =
\frac{1}{i\omega} \int\! d z' \partial_z g(z,z',\omega)E^{ext}(\omega)  \, .     
                                                              \label{e3}
\end{equation}
Since in the absence of the barrier no charge is induced in the leads, only the 
translationally non--invariant part of the conductance, $g_b (z,z',\omega)$, 
contributes to $\rho^{(0)}$. 
It is convenient to employ   Fourier representation. 
Then, using  Eq.~(\ref{n1}), one may  
rewrite Eqs.~(\ref{e2}), (\ref{e3}) in the following form 
\begin{eqnarray}
&&\left[  \frac{i \omega S}{4\pi} - g_0(q,\omega) \right] E^{ind}(q,\omega) - 
                                                           \label{e5}\\
&&\int\! {d q'\over 2\pi}\,  g_b(q,q',\omega) E^{ind}(q',\omega) = 
g_b(q,0,\omega)E^{ext}(\omega)  \, , 
                                                           \nonumber
\end{eqnarray}
where e.g. 
\begin{equation} 
g_b(q,q',\omega) = 
\int\!\! \int\!\! dz dz'\, e^{iqz} g_b(z,z',\omega) e^{-iq'z'}\, . 
                                                              \label{e4a}
\end{equation}

Once Eq.~(\ref{e5}) is solved, the induced voltage drop across the barrier is 
given by $V=\int dz E^{ind}(z,\omega)$, where the limits of integration are taken 
to be much larger than any microscopic scale of the problem (see below), but still 
much smaller than the length, ${\cal L}$, of the ring which encloses the driving 
time--dependent magnetic flux. In the Fourier representation one obtains
\begin{equation} 
V =  E^{ind}(q\to 0,\omega) \, ,
                                                              \label{e6}
\end{equation}  
where $q\to 0$ in such a way that ${\cal L}^{-1}\ll q_0 \ll \omega/v_F$. 
The divergence-less part of the current, $I_0$, originates from the convolution 
of $g_0(z-z')$ and $E^{ext}$: 
\begin{equation} 
I_0= g_0(q=0,\omega)  E^{ext}(\omega)  \, .  
                                                              \label{e7}
\end{equation}  
All other terms in Eq.~(\ref{e1a}) 
describe currents localized near the barrier. 
As a result the four--terminal conductance, $\tilde g =I_0/V$, may be 
expressed as 
\begin{equation} 
\tilde g =  g_0(0,\omega) \frac{E^{ext}(\omega)}{E^{ind}(q\to 0,\omega)} 
                                                              \label{e8}
\end{equation}
We shall proceed now by  calculating the non--local conductance, 
$g(z,z',\omega)$, 
and then solving Eq.~(\ref{e2}) for the various cases of practical interest.

\section{Barrier with Long Clean Leads}
\label{s2}

If there is no scattering of electrons in the leads,
the Green's functions outside the barrier region 
may be  expressed as \cite{foot2} 
\widetext
\top{-2.8cm} 
\begin{equation} 
G^{R}_{ij}(\epsilon;z,z')= 
\frac{-im}{\sqrt{p_{i}p_{j}}} 
\left\{ \begin{array}{ll}
t_{ij}(\epsilon) e^{i p_{i}|z|+ i p_{j}|z'|}\, , &
zz' < 0\,;  \\
 \delta_{ij} e^{i p_i |z-z'|}     + 
r_{ij}(\epsilon) e^{i p_{i}|z|+i p_{j}|z'|}\, ,
\,\,   & zz' >0 \, , 
\end{array} \right. 
                                                              \label{w6}
\end{equation}  
where $p_i^2/(2m) =mv_i^2/2= \epsilon-\epsilon_i$, {\em etc}. 
The transmission and  reflection matrices of the barrier, 
$t_{ij}(\epsilon)$ and $r_{ij}(\epsilon)$, obey the unitarity condition  
\begin{equation} 
\sum\limits_{j} \left[ 
|t_{ij}(\epsilon)|^2+ |r_{ij}(\epsilon)|^2 
\right]  = 1\, .
                                                              \label{w7}
\end{equation} 
Substituting Eq.~(\ref{w6}) into Eq.~(\ref{w2}), one can perform the energy 
integration assuming that $t_{ij}(\epsilon)$ and 
$r_{ij}(\epsilon) $ are slowly varying functions of energy on the scales 
of both 
temperature and frequency, $\omega$. As a result one obtains for the 
non--local conductance  outside the barrier 
\begin{equation} 
g(z,z',\omega)=  \frac{e^2}{\pi\hbar}
\sum\limits_{ij} 
\left\{ \begin{array}{ll}
t_{ij} (t_{ij})^*
e^{i\omega(|z|/v_i+|z'|/v_j)}\, , & zz'<0 \, ; \\ 
\delta_{ij} e^{i\omega |z-z'|/v_i} -
r_{ij} (r_{ij})^*  e^{i\omega( |z|/v_i + |z'|/v_j)}\, , \,\,\, &zz'>0 \, ,
\end{array} \right. 
                                                              \label{w8}
\end{equation} 
\bottom{-2.7cm}
\narrowtext
\noindent  
%\hskip -.3cm 
where $t_{ij}=t_{ij}(\epsilon_F)$ and $\epsilon_F$ is the Fermi energy. 
Eq. (\ref{w8}) is the zeroth order term 
in the small parameter $\omega/\epsilon_F$. Note, however, that even for 
small $\omega$ the combination  
$\omega z/v_i$ is not necessarily  small. 
 
Setting $\omega=0$ in Eq.~(\ref{w8}) leads to
\begin{equation} 
g\equiv g(z,z',\omega=0) =  \frac{e^2}{\pi\hbar}\, 
\Tr\,  t t^{\dagger}\, .
                                                              \label{w9a}
\end{equation}
The fact that at $\omega=0$ the conductance kernel is coordinate independent 
is a manifestation of the continuity equation. As a result the total current 
is divergence-less and given by $I_0=g\int dz E(z)$. The integral on the r.h.s. 
is a total voltage drop across the entire wire (and {\em not} just across the 
barrier). Therefore, the quantity $g$ defined by Eq.~(\ref{w9a}) represents the 
{\em two--terminal} conductance in agreement with the Landauer expression, 
Eq.~(\ref{q1}). To analyze the four--terminal setup, one has to solve 
the self-consistent equation  Eq.~(\ref{e2}) and thus has to keep frequency 
the dependence of Eq.~(\ref{w8}).    

For simplicity we restrict ourself to the case of 
no channel mixing inside the barrier region, i.e. 
$t_{ij}= t_i \delta_{ij}$ and $r_{ij}= r_i \delta_{ij}$, where, according to 
Eq.~(\ref{w7}), $|t_i|^2+|r_i|^2=1$. In this case Eq.~(\ref{w8}) may be
simplified further (see Ref. \cite{Economou81}):
\begin{equation} 
g(z,z',\omega)=  \frac{e^2}{\pi\hbar}
\sum\limits_{i}\left[
e^{i \omega |z-z'|/v_i}  -
|r_{i}|^2 
e^{i\omega (|z|+|z'|)/v_i} \right]  \, . 
                                                              \label{w9}
\end{equation}
The two terms on the right hand side correspond to 
$g_0$ and $g_b$ introduced in Eq.~(\ref{n1}). In the Fourier representation 
these terms take  the form 
\begin{eqnarray}
&&g_0(q,\omega) = - \frac{e^2}{\pi\hbar} 
\sum\limits_{i} 
{2v_i i\omega  \over v_i^2 q^2 - \omega^2} \, ;
                                                              \label{w12}\\
&&g_b(q,q',\omega) = - \frac{e^2}{\pi\hbar} 
\sum\limits_{i}  |r_{i}|^2
{2v_i i\omega \over v_i^2 q^2 - \omega^2}\,
{2v_i i\omega \over v_i^2 q'^2 - \omega^2} \, . 
                                                              \label{w13}
\end{eqnarray}
These results may be derived directly in the momentum representation starting 
from the Kubo formula and employing the following expression for the Green's 
functions 
\begin{eqnarray} 
&&G^{R}_{jj}(\epsilon;p,p') = 
G^{R}_{j}(\epsilon;p) 2\pi\delta_{p,p'} +  
                                                   \label{n2}\\
&&i v_j G^{R}_{j}(\epsilon;p) G^{R}_{j}(\epsilon;p') 
\left\{ 
\frac{t_j-1 + r_j}{2} + \frac{p p'}{p_j^2}\, \frac{t_j-1 - r_j}{2} 
\right\}\, .
                                                  \nonumber
\end{eqnarray}
Here $G^{R}_{j}(\epsilon;p)=(\epsilon-\epsilon_j - p^2/(2m)+i\eta)^{-1}$ 
is the retarded Green's function of the $j-th$ channel in the wire without 
a barrier (see Eq.~(\ref{w6})). 

By introducing the dimensionless momentum $l=q v_F/\omega$ and normalized
induced electric field 
${\tilde E}(l)\equiv \omega E^{ind}(l\omega/v_F,\omega)/(v_FE^{ext}(\omega))$,
Eq.~(\ref{e5}) acquires the form 
\begin{equation}
\left[ 2i \frac{\omega^2}{\tilde\omega_p^2} - g_0(l) \right] {\tilde E}(l) - 
\int\! {d l'\over 2\pi} \,  g_b(l,l') {\tilde E}(l')= 
g_b(l,0) \, ,     
                                                              \label{e5a}
\end{equation}
where 
$\tilde\omega_p^2 \equiv 8 e^2v_F/(\hbar S)=\omega_p^2 3\pi\hbar^2/(k_F^2 S)$, 
with 
$\omega_p^2=4\pi e^2n/m$ being the plasma frequency. We have introduced 
the rescaled (frequency independent) conductance as 
\begin{eqnarray}
&&g_0(l) \equiv -i
\sum\limits_{i} {2\tilde v_i  \over \tilde v_i^2 l^2 - 1} \, ;
                                                              \label{w12a}\\
&&g_b(l,l') \equiv 
\sum\limits_{i} |r_{i}|^2 {2\tilde v_i\over \tilde v_i^2 l^2 - 1}
\, {2\tilde v_i\over \tilde v_i^2 (l')^2 - 1}  \, , 
                                                              \label{w13a}
\end{eqnarray}
where $\tilde v_i \equiv v_i/v_F$.
At small frequency, 
$\omega\ll \tilde\omega_p$, one may neglect the first term on the l.h.s. of 
Eq.~(\ref{e5a}). The remaining equation is manifestly frequency independent. 
We shall, however, keep the frequency dependent term for the time being since it 
will allow us to discuss the characteristic length scales of the problem. 
We now proceed to solve Eq.~(\ref{e5a}) in some particular cases.

\subsection{Single--channel case}
\label{s31}

First we apply the formalism to the well--known single--channel case. 
In this case the integral equation (\ref{e5a}) has a separable 
kernel and may be easily solved. 
We neglect 
for a moment the frequency dependent term. Substituting 
Eqs.~(\ref{w12a}) and (\ref{w13a}) with $\tilde v_1=1$ 
into Eq.~(\ref{e5a}), one obtains 
\begin{equation}
\frac{{\tilde E}(l)}{2i|r|^2} + 
\int\! {d l'\over 2\pi}\,  \frac{{\tilde E}(l')}{(l')^2 - 1} = 1 \, .     
                                                              \label{eA1}
\end{equation} 
The elementary solution of this equation is 
\begin{equation}
{\tilde E}(l) = 2i \frac{|r|^2}{1-|r|^2} \, .     
                                                              \label{eA2}
\end{equation} 
Employing Eq.~(\ref{e8}), one immediately obtains the famous Landauer 
expression, Eq.~(\ref{q2}). This agrees with the conclusions  
obtained by Thouless \cite{Thouless81} by imposing  strict 
charge neutrality outside the barrier. Indeed, the fact that  
$E^{ind}(q)=\mbox{const}$ means that $E^{ind}(z)\propto \delta(z)$. 
Thus, in this approximation 
there is no induced electric field and no induced charge density
outside the barrier. To recover the shape of the charge distribution one must
keep the frequency dependent term in Eq.~(\ref{e5a}). An elementary calculation 
gives 
\begin{equation}
\rho(z,\omega)\propto \mbox{sign}(z)
\exp \left\{ - \frac{|z|\tilde \omega_p}{v_F}
\sqrt{1-\frac{\omega^2}{\tilde\omega_p^2}} \right\}\, , 
                                                              \label{eA3}
\end{equation} 
showing that  entire charge redistribution is confined to 
within the screening length, $\kappa^{-1}\equiv v_F/\tilde\omega_p$, 
near the barrier. Since we keep only the long wave--length components of the 
conductance kernel, Eq.~(\ref{eA3}) is valid  only if the inequality 
$\kappa < k_F$ is fulfilled (if not there are oscillations on the scale 
$(2k_F)^{-1}$). Although the initial set of equations,  
(\ref{e1a})--(\ref{e1c}), included the characteristic length 
$v_F/\omega $, the expression (\ref{eA3}) depends only on the much smaller scale, 
$\kappa^{-1}$. This is a particular property of the 
single--channel   case only. In  the multi--channel case
the length scale $v_F/\omega $ does not drop out of the solution of 
the self--consistent equations. This demands  separate treatments for wires 
whose lengths are  larger or smaller than $v_F/\omega $.

If one keeps the frequency dependent term in Eq.~(\ref{e5}), the 
resulting four--terminal conductance is 
\begin{equation}
\tilde g= \frac{e^2}{\pi\hbar} 
\left[ 
\frac{|t|^2}{|r|^2} \left(1-\frac{\omega^2}{\tilde \omega_p^2} \right) -  
i\frac{\omega}{\tilde\omega_p}\sqrt{1-\frac{\omega^2}{\tilde\omega_p^2}} 
\right]     \, .     
                                                              \label{eA4}
\end{equation} 
At small frequency this expression describes the Landauer resistance in parallel 
with an effective  capacitance 
\begin{equation}
C_0\equiv \frac{S}{4\pi(2\kappa^{-1}) }   \, ,     
                                                              \label{eA5}
\end{equation} 
which is the classical result for a plane capacitor of the area $S$ and
spacing $2\kappa^{-1}$.  In our approximations the screening 
length is assumed to be much larger than both the wavelength and the barrier 
width. If this is not the case, one has to keep the next order in 
$\omega/\epsilon_F$ as well as short wave--lengths in the expression for 
the conductance kernel, to obtain the effective capacitance.

\subsection{Equal velocity channels}
\label{s32}

Another exactly solvable case, which will be used in Sec. \ref{s7} is that  
of $N$ channels having the same velocity, 
$\tilde v_i=1\,;\,\, i=1\ldots N$. 
Calculations exactly parallel to the previous case lead to 
\begin{equation}
\tilde g = \frac{e^2}{\pi\hbar}\, \frac{N \sum\limits_i^N|t_i|^2}
{\sum\limits_i^N | r_i|^2 } \, .
                                                              \label{eB1}
\end{equation} 
This result is in agreement with BILP, Eq.~(\ref{q3}).
If only $M\ll N$ channels are open, 
$t_i = 0, |r_i|=1$ for  $i=M+1\ldots N$, Eq.~(\ref{eB1}) simplifies to 
\begin{equation}
\tilde g= \frac{e^2}{\pi\hbar}\sum\limits_i^M|t_i|^2 + 
O\left(\frac{M}{N}\right)\, .
                                                               \label{eB2}
\end{equation}

\subsection{Weakly reflecting channels}
\label{s34}

The case of a weakly reflecting barrier, $|r_i|\ll 1\,; \,\, i=1\ldots N$, 
may be considered by treating $g_b(l,l')$ as a perturbation in 
Eq.~(\ref{e5a}). Employing obvious operator notations 
and omitting the frequency dependent term, one may write the formal solution of 
Eq.~(\ref{e5a}) as 
\begin{eqnarray}
{\tilde E}(l) &&= 
- (\hat g_0 + \hat g_b)^{-1}\hat g_b|\delta \rangle
                                                      \label{eD1}\\
&&=(-\hat g_0^{-1}\hat g_b + \hat g_0^{-1}\hat g_b\hat g_0^{-1}\hat g_b -
\dots)|\delta\rangle \, , 
                                                      \nonumber
\end{eqnarray} 
where $|\delta\rangle = (1, 0, 0,\ldots)$, where the first entry refers to $l=0$ 
component. Since  
$\hat g_0$ is diagonal and may be easily inverted, all terms in the expansion, 
Eq.~(\ref{eD1}), may be written in quadratures. We restrict ourself to 
the leading term only. In this approximation one obtains 
\begin{equation}
{\tilde E}(l) = i\,\, 
\frac{\sum\limits_i 4 \tilde v_i^2 |r_i|^2 (\tilde v_i^2 l^2 - 1)^{-1} }
{\sum\limits_i 2\tilde v_i  (\tilde v_i^2 l^2 - 1)^{-1} } \, . 
                                                      \label{eD2}\\
\end{equation}  
Substitution of Eq.~(\ref{eD2}) into Eq.~(\ref{e8}) leads to the 
result, announced in Sec. \ref{s1}, 
\begin{equation} 
\tilde g=\frac{e^2}{\pi\hbar}\,  
\frac{\left[ \sum\limits_i v_i \right]^2 }
{\sum\limits_i v_i^{2} |r_i|^2}\, .
                                                       \label{eD3}
\end{equation}  
Note that, apart from 
the single--channel case, the approximate solution, Eq.~(\ref{eD2}), 
depends on the momentum, $l$, and hence on the $z$--coordinate. This means 
that  strict charge neutrality does not occur in 
general. Rather, the resulting charge density exhibits a
spatial modulation on the scale $v_F/\omega$.

\subsection{Weakly transmitting  channels}
\label{s35}

We turn now to the high barrier case, $|t_i|\ll 1\,; \,\, i=1\ldots N$. 
Employing $|r_i|^2 = 1 - |t_i|^2$, we rearrange Eqs.~(\ref{w12a}), (\ref{w13a})
 as 
\begin{equation} 
g(l,l')=  g_1(l,l') + g_t(l,l')\, ,
                                                              \label{eE1}
\end{equation}
where
\begin{eqnarray}
&&g_1(l,l') = 
\sum\limits_{i} {2\tilde v_i \over \tilde v_i^2 l^2 - 1} 
\left[ 
-2\pi i \delta_{l,l'}  +  {2\tilde v_i \over \tilde v_i^2 (l')^2 - 1}
\right] \, ;
                                                              \label{eE2}\\
&&g_t(l,l') = -
\sum\limits_{i} |t_{i}|^2\, 
{2\tilde v_i \over \tilde v_i^2 l^2 - 1}
\, {2\tilde v_i \over \tilde v_i^2 (l')^2 - 1}  \, . 
                                                              \label{eE3}
\end{eqnarray}
Here $g_1$ describes two disconnected half--wires, and $g_t$ -- 
the perturbation due to the small tunneling transparency of the barrier. 
Since the conductance of the infinite barrier is zero, one expects that 
the operator $\hat g_1$ has a zero eigenvalue. Indeed, it is easy to check that 
\begin{equation} 
\hat g_1|1 \rangle \equiv \int\!\! {dl'\over 2\pi}  g_1(l,l') = 0 \, ,
                                                              \label{eE4}
\end{equation}  
where $|1 \rangle$ is the abstract vector in the momentum space given by 
$(1,1,1 \ldots)$, whose entries refer to different values of $l$.  
As a result, the operator  $\hat g_1$ is not invertible, which complicates 
the perturbation theory. To overcome this difficulty we pass 
to the basis of eigenstates of the operator $\hat g_1$, 
\begin{equation} 
\hat g_1 |a\rangle = \lambda_a |a\rangle  \, ,
                                                              \label{eE5}
\end{equation} 
where $a=1,2,\ldots$ and $\lambda_1 =0$. 
In this basis Eq.~(\ref{e5a}) takes the form 
\begin{eqnarray}
&&(\hat g_t)_{1a'}{\tilde E}_{a'} = - (g_b)_1\, ;
                                                        \label{eE6}\\
&&\lambda_a {\tilde E}_a + 
(\hat g_t)_{aa'}{\tilde E}_{a'} = -(g_b)_a \, , 
\hskip 1.5cm a \neq 1\, , 
                                                              \label{eE7}
\end{eqnarray}
where $(g_b)_a = \langle a| \hat g_b |\delta\rangle$, 
and summation over repeated indexes is assumed.    
Taking advantage of the smallness of $\hat g_t\propto |t_i|^2$, one may solve 
these equations iteratively. 
Neglecting  first $\hat g_t$ on the l.h.s. of  
Eq.~(\ref{eE7}), one obtains in the zeroth approximation 
\begin{eqnarray} 
&&{\tilde E}_a = -\frac{(g_b)_a}{\lambda_a }\,, 
\hskip 1.5cm a \neq 1\, ;
                                                              \label{eE8}\\
&&{\tilde E}_1=-\frac{(g_b)_1}{(\hat g_t)_{11} } + 
\sum\limits_{a'\neq 1} 
\frac{(\hat g_t)_{1a'} }{(\hat g_t)_{11} } \,  
\frac{(g_b)_{a'}}{\lambda_{a'}}\, .
                                                              \label{eE9} 
\end{eqnarray} 
Repeatedly substituting the solution back to Eq.~(\ref{eE7}), one may 
obtain the higher order terms. 
In the leading (minus first) order in $\hat g_t$ the solution has the form 
\begin{equation} 
{\tilde E}_a =- \frac{(g_b)_1}{(\hat g_t)_{11} }\, 
\delta_{a,1}\, .
                                                              \label{eE10}
\end{equation}  
To this order of accuracy on has to put $|r_i|^2 = 1$ in the numerator and 
therefore $(g_b)_1 =  - g_0(0,\omega)$. 
By virtue of Eq.~(\ref{eE4})  $(\hat g_t)_{11} = (\hat g)_{11}$. 
After substitution into  Eq.~(\ref{e8}) one obtains for the 
four--terminal conductance 
\begin{equation} 
\tilde g = (\hat g)_{11}\, . 
                                                              \label{eE11}
\end{equation}  
This expression includes the exactly known $|1\rangle$ eigenfunction 
only. A straightforward calculation of $(\hat g)_{11} $  
gives  the 
result, equivalent to the two--terminal expression,
\begin{equation} 
\tilde g=\frac{e^2}{\pi\hbar}
\sum\limits_{i} |t_{i}|^2\, .
                                                              \label{eE12}
\end{equation}  
This is the  expected result, since for a high barrier, two and 
four--terminal measurements should yield the same result.  
Note, that to this order ${\tilde E}(l)=\mbox{const}$, which is consistent 
with  strict charge neutrality. To evaluate corrections to the leading order, 
Eq.~(\ref{eE11}), one needs an explicit form of the eigenfunctions of the 
$\hat g_1$ operator, Eq.~(\ref{eE5}), which is in general a hard problem.

\section{Barrier with Disordered Leads}
\label{s5}

\begin{figure}
\vglue 0.5cm
\hspace{0.01\hsize}
\epsfxsize=0.9\hsize
\epsffile{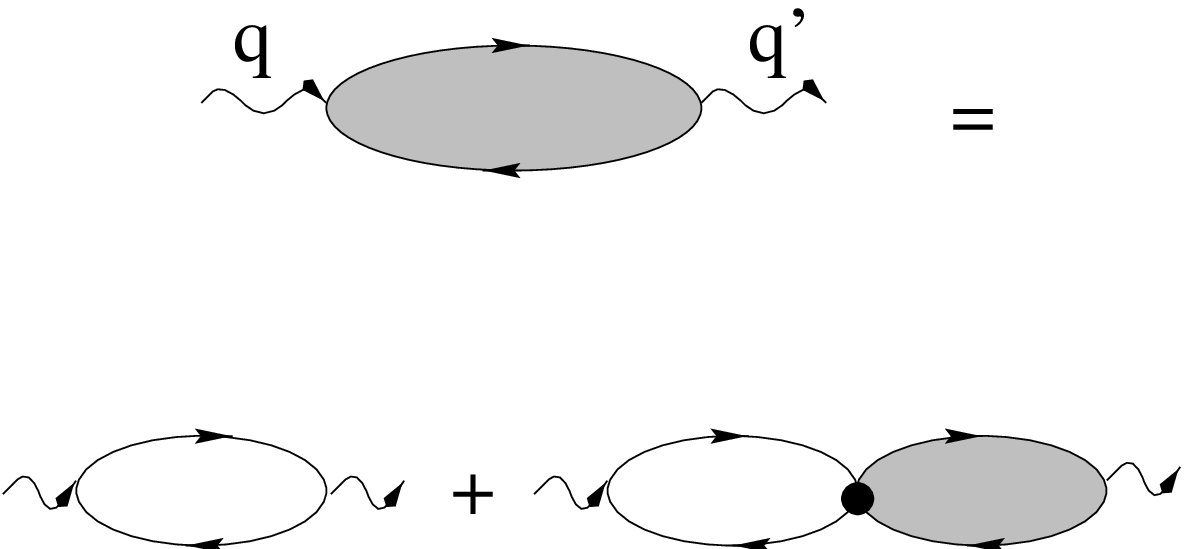}
\vspace{0.1\hsize}
\refstepcounter{figure} \label{fig0}
{\small FIG.\ \ref{fig0} Diagrammatic representation of  
the dynamic compressibility $\Pi(q,q',\omega)$, denoted by the shaded bubble; 
the quantity $\eta(q,q',\omega)$ is denoted by the empty bubble. 
The full lines represent electron Green's 
function, Eqs.~(\ref{n2}), (\ref{n4}) and the full dot 
represents a point--like elastic
scatterer with the amplitude $(\tau\nu)^{-1}$.   
 \par}
\end{figure}

We consider now the case, where the leads attached to the barrier contain 
weak elastic disorder. We restrict ourselves to uncorrelated short--range 
scatterers, with  mean free path $l^{el}$ and mean free time $\tau$ 
related by $l^{el} = v_F \tau$.
The leads are assumed to be much longer than the mean free path: $L\gg l^{el}$.
As a result, the momentum distribution function far from the barrier is 
expected to be isotropic (spherical). If a current passes through the wire,
the distribution function is a shifted Fermi--sphere. Our immediate 
goal is to calculate the disorder--averaged non--local conductance of such 
a system (leads with the barrier), substitute it to Eq.~(\ref{e5}) and 
solve the latter.
To this end one has to calculate the average product of 
two Green's functions, Eq.~(\ref{w2}). The average single particle  
Green's function of the leads with the barrier is given by  Eq.~(\ref{n2})
where, in the disordered case \cite{Doniach}, 
\begin{equation} 
G^{R}_{j}(\epsilon;p)=\frac{1}{\epsilon-\epsilon_j - p^2/(2m)+
i/(2\tau) }\, . 
                                                     \label{n4}
\end{equation} 
Employing the continuity relation, one may express   
the conductance kernel as
\begin{equation} 
g(q,q',\omega) = -\frac{2 e^2 i\omega}{ q q'} 
\left[
2\pi\delta_{q,q'}\sum\limits_i \frac{1}{\pi v_i} + 
i\omega \Pi(q,q',\omega) 
\right]\, .
                                                     \label{n5}
\end{equation} 
The two terms in the square brackets represent the static and dynamic 
parts of the compressibility. In the diffusive approximation 
($k_F l^{el} >> 1$) the second term is given by the sequence of the bubble  
diagrams \cite{Doniach}, depicted in Fig. \ref{fig0}. 
Therefore the quantity $\Pi(q,q',\omega)$ 
may be found as a solution of the following integral equation 
\begin{equation} 
\Pi(q,q') = \eta(q,q') + 
\frac{1}{\tau\sum\limits_i  (\pi v_i)^{-1} }
\int\!\!{dq''\over 2\pi} \eta(q,q'') \Pi(q'',q')\, . 
                                                     \label{n6}
\end{equation} 
Here $\tau\sum\limits_i (\pi v_i)^{-1}$ is the inverse 
scattering amplitude 
and $\eta(q,q',\omega)$  is the single bubble (see Fig. \ref{fig0})
\begin{equation} 
\eta(q,q',\omega) ={1\over 2\pi}
\sum\limits_{p,p',j} G^{R}_{jj}(\epsilon_+;p_+,p'_+)\,
G^{A}_{jj}(\epsilon_-;p'_-,p_-)\, ,
                                                     \label{n7}
\end{equation}
where $p_{\pm} = p\pm q/2$, $p'_{\pm} = p'\pm q'/2$ and the average Green's 
functions are given by Eqs.~(\ref{n2}) and (\ref{n4}). 
Performing momentum summations, one 
obtains in the long wavelength limit 
($q,q'\ll k_F$, $\omega\ll \epsilon_F$) 
\begin{eqnarray} 
&&\eta(q,q',\omega) =
\sum\limits_{i}\frac{\tau}{\pi v_i} 
\frac{1-i\omega\tau}{(1-i\omega\tau)^2+(v_iq\tau)^2} 
2\pi \delta_{q,q'} + 
                                              \label{n8}\\
&&{1\over 2\pi }
\sum\limits_{i}\frac{ |r_i|^2 4v_i^2  q q' \tau^4} 
{[(1-i\omega\tau)^2+(v_iq\tau)^2] [(1-i\omega\tau)^2+(v_iq'\tau)^2] } \, .
                                                    \nonumber  
\end{eqnarray} 

In the clean limit, $\omega\tau\ \gg 1$, the sequence is 
dominated by the single bubble resulting in $\Pi=\eta$. Under these conditions 
Eqs.~(\ref{n5}) and (\ref{n8}) lead to the previous result given by 
Eqs.~(\ref{w12}), (\ref{w13}). In the following we concentrate on the opposite, 
diffusive limit, $v_Fq\tau \ll 1$,  $\omega\tau\ \ll 1$. 

In the absence of the barrier, $r_i =0$, and 
Eq.~(\ref{n6}) may be easily solved, resulting in 
\begin{equation} 
\Pi(q,q',\omega) = \nu\, \frac{2\pi \delta_{q,q'}}{D q^2 -i\omega} 
\equiv \Pi_0(q,\omega)\, 2\pi \delta_{q,q'}  \, , 
                                                     \label{n9}
\end{equation}
where the single particle density of states (per spin), $\nu$, 
and the diffusion constant, $D$, are defined as 
\begin{equation} 
\nu = \sum\limits_i {1\over \pi v_i}\, ; \,\,\,\,
D = \frac{\tau \sum\limits_i v_i}{\sum\limits_i v^{-1}_i} \, .
                                                     \label{n10}
\end{equation}
Substituting Eq.~(\ref{n9})
into Eq.~(\ref{n5}), one obtains the standard diffusive expression for the 
conductance of a disordered wire 
\begin{equation} 
g_0(q,\omega) = - {e^2\over \pi \hbar} \sum\limits_i 
\frac{2v_i i\omega \tau }{Dq^2 -i\omega}\, . 
                                                     \label{n11}
\end{equation}
Next we shall look for $g_b(q,q',\omega)$ and then for solutions of the 
self--consistent equation (\ref{e5}) in some particular cases.

\subsection{Equal velocity channels}
\label{s51}

In the case of $N$ channels with the same velocity, $v_i= v,\, i=1\dots N$, 
the integral equation (\ref{n6}) has a separable 
kernel and thus may be solved exactly. After  simple algebra one obtains for 
the barrier--induced part of the conductance 
\begin{equation} 
g_b(q,q',\omega) = 
\frac{- e^2/(\pi \hbar) \sum_j |r_j|^2} 
{\sum_j \left( |t_j|^2 -i|r_j|^2 \sqrt{i\omega\tau} \right) } 
\frac{ 2v i\omega\tau }{Dq^2 -i\omega}  
\frac{ 2v i\omega\tau }{Dq'^2 -i\omega} \, . 
                                                     \label{nB1}
\end{equation}
This expression is valid for $\omega\tau \ll 1$; however, we have retained 
the frequency dependent term in the denominator since the condition 
$\sqrt{\omega\tau} \ll |t_j|^2/|r_j|^2$ is not assumed. 
Substituting Eqs.~(\ref{n11}), 
(\ref{nB1}) into the self--consistent equation for the induced electric field,
Eq.~(\ref{e5}),  and solving the latter, one obtains, for 
$\omega\ll \tilde \omega_p$,
\begin{equation} 
E^{ind}(q) = 2v\tau 
\frac{\sum_j |r_j|^2} {\sum_j|t_j|^2}\, E^{ext}\, . 
                                                     \label{nB2}
\end{equation}
Finally, employing Eqs.~(\ref{e8}) and (\ref{n11}), yields
the four--terminal conductance 
\begin{equation}
\tilde g = \frac{e^2}{\pi\hbar}\, \frac{N \sum\limits_j^N|t_j|^2}
{\sum\limits_j^N | r_j|^2 } \, .
                                                              \label{nB3}
\end{equation} 
in agreement with a clean case, cf. Eq.~(\ref{eB1}).
Although for the sake of compactness we wrote all intermediate expressions
for the small frequency limit, $\omega\tau \ll 1$, one may actually perform all 
the calculations and obtain  
the final result, Eq.~(\ref{nB3}), for any $\omega\tau$  (the only 
limitation is $\omega\ll\tilde \omega_p$).

\subsection{Weakly reflecting channels}
\label{s52}

One may formally solve the integral equation (\ref{n6}) employing 
perturbation theory with the small parameter,  $|r_i|^2 \ll 1$. 
The first order correction  to $\Pi(q,q',\omega)$ for $\omega\tau \ll 1$ 
is  
\begin{equation} 
\delta \Pi (q,q',\omega) = {\tau^2\over 2\pi} 
\sum\limits_i |r_i|^2 
\frac{2v_i q}{Dq^2 -i\omega}  \frac{2 v_i q'}{Dq'^2 -i\omega} \, . 
                                                     \label{n12}
\end{equation}
Employing  Eq.~(\ref{n5}), one obtains in the same order 
\begin{equation} 
g_b(q,q',\omega) = - {e^2\over \pi \hbar} 
\sum\limits_i |r_i|^2 
\frac{2 v_i i\omega\tau }{Dq^2 -i\omega}  
\frac{2 v_i i\omega\tau }{Dq'^2 -i\omega} \, . 
                                                     \label{n13}
\end{equation}
Note that Eqs.~(\ref{n11}), (\ref{n13}) are almost exact analogs of 
Eqs.~(\ref{w12}), (\ref{w13}), where the ballistic propagators are 
replaced by  diffusive ones. The important difference, however, 
is that the validity of Eqs.~(\ref{w12}), (\ref{w13}) is not restricted to 
small reflection coefficients, $r_i$. On the other hand, in the diffusive 
problem an electron can bounce between the barrier and impurities, 
and thus the exact conductance contains  higher powers of 
$|r_i|^2$. Eqs.~(\ref{n11}) and  (\ref{n13}) are only the first two terms in 
the infinite series. 

We now substitute  Eqs.~(\ref{n11}), (\ref{n13}) into Eq.~(\ref{e5}) and solve 
the integral equation. Since the kernel, Eq.~(\ref{n13}), is separable, its 
solution is elementary.  The result for the induced electric field in 
 the zero frequency limit is 
\begin{equation} 
E^{ind}(q) = 
\frac{\tau \sum_i 2v_i^2|r_i|^2}{\sum_i v_i}\,  E^{ext}\, .
                                                     \label{n14}
\end{equation}
Employing Eq.~(\ref{e8}) and (\ref{n11}), one obtains the same result for 
the four--terminals 
conductance, $\tilde g$, as in the clean case -- Eq.~(\ref{q4}). 
This statement is actually valid for any value of $\omega\tau$, 
provided $|r_i|^2 \ll 1$ and $\omega\ll \tilde\omega_p$.

\subsection{Weakly transmitting  channels}
\label{s53}

For a completely reflecting  barrier, $|r_i|= 1$, 
one may easily check the following equality 
\begin{equation} 
\int\! {dq'\over 2\pi} \left[2\pi\delta_{q,q'}- 
\frac{1-i\omega\tau}{\nu\tau}\eta^{|r_i|= 1}(q,q',\omega)
\right] \frac{1}{q'} = 0 \, .
                                                     \label{nC0}
\end{equation}
With its help and employing Eqs.~(\ref{n5}), (\ref{n6}) and 
(\ref{n8}) with $|r_i|^2 =1 $, one may prove that 
\begin{equation} 
\int\! {dq'\over 2\pi} g^{|r_i|= 1} (q,q',\omega) = 0 \, 
                                                     \label{nC1}
\end{equation}
for any $q$ and $\omega$. This identity simply reflects the fact that 
an electric field localized under a  completely reflecting barrier cannot 
induce any current.  
Thus, to solve the self--consistent field equation for $|t_i| \ll 1$
one has to invert an operator having one almost vanishing eigenvalue. 
Following the method described in section  \ref{s35}, one obtains for the 
the four--terminal conductance  
\begin{equation} 
\tilde g = (\hat g)_{11} \, ,
                                                              \label{nc2}
\end{equation}  
where $(\hat g)_{11} = (2\pi)^{-2}\int dq dq' g(q,q')$.  Calculation of this 
matrix element is not as straightforward as in the clean case. 
To compute it one has first to solve  the integral equation (\ref{n6}). 
Simple algebra reduces that problem to the inversion of the operator 
which differs by a small factor $(\sim |t_i|^2)$ 
from the one written in the square brackets 
on the left hand side of Eq.~(\ref{nC0}). According to Eq.~(\ref{nC0}) 
the latter has exactly zero eigenvalue, with the corresponding eigenvector 
$\propto 1/q'$.
One has to employ once again the method described in section  \ref{s35} to 
invert an  operator with one small eigenvalue. As a result of this procedure 
one obtains Eq.~(\ref{eE12}), the same as in  clean case.

\section{Barrier in a Ballistic Constriction}
\label{s7}

\begin{figure}
\vglue 0.5cm
\hspace{0.01\hsize}
\epsfxsize=0.9\hsize
\epsffile{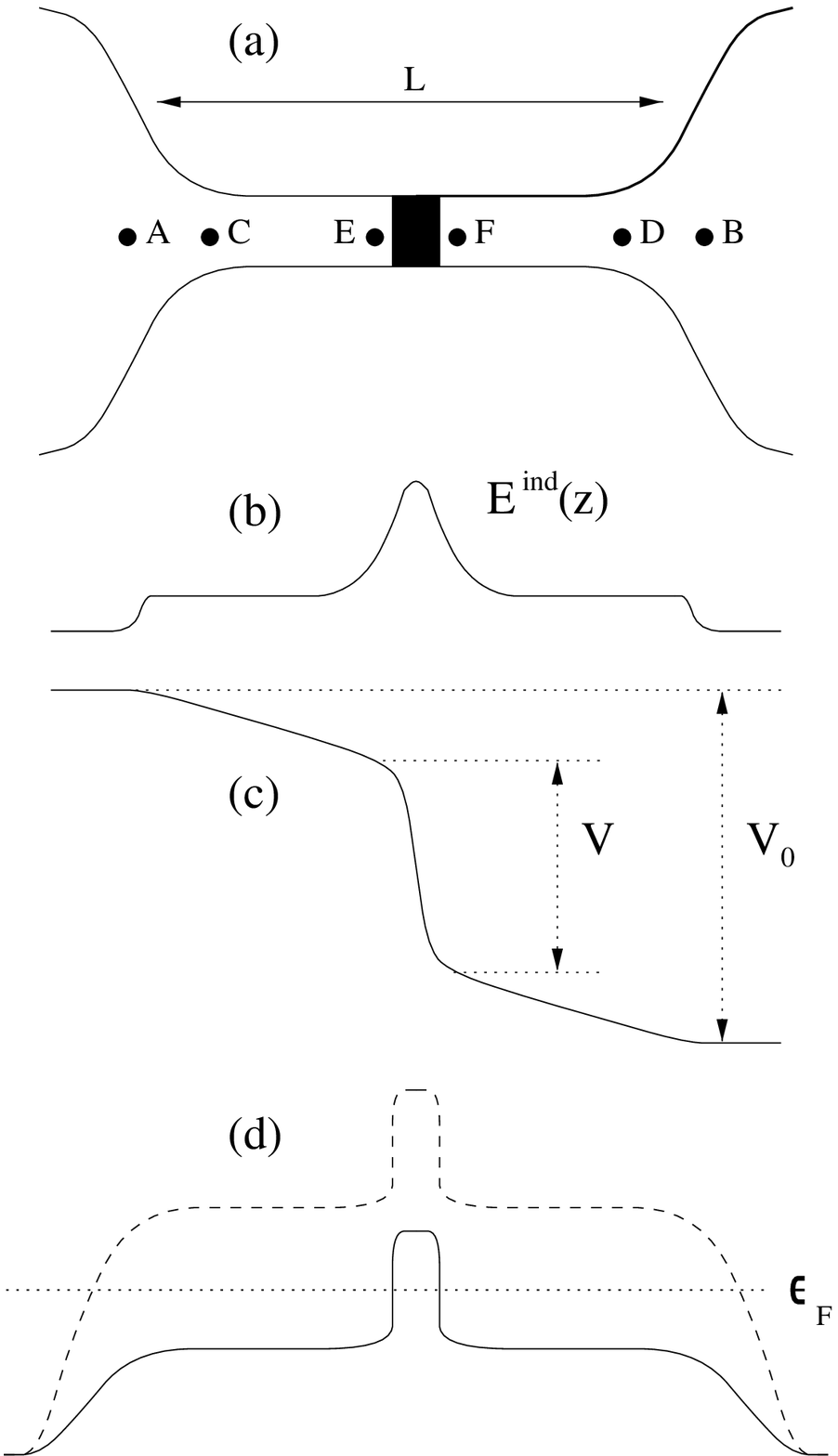}
\refstepcounter{figure} \label{fig2}
\vspace{0.1\hsize}
{\small FIG.\ \ref{fig2} (a) Schematic drawing of the adiabatic constriction with 
the barrier. (b) The induced electric field, $E^{ind}(z)$, given by 
Eqs.~(\ref{yy7}), (\ref{yy7a}). (c) The scalar potential which corresponds to the 
induced electric field, $E^{ind}(z)$. (d)
The effective potential $\epsilon_i(z) +V_b(z)$ for the 
open (full line) and closed (dashed line) channels. The dotted line represents 
the  Fermi energy, $\epsilon_F$.  
 \par}
\end{figure}

We now consider a barrier embedded into a ballistic 
adiabatic {\em constriction} between the two wide reservoirs 
(see Fig. \ref{fig2}a). 
We shall
assume that the length of the constriction, $L$, 
satisfies the conditions 
$L < v_F/\omega$ and $L< l^{el}$. This is the geometry considered by  
BILP 
\cite{Buttiker85} and others (see Ref. \cite{Stone89} and references therein).
Since  the cross-sectional area of the constriction, $S(z)$, is a smooth 
function of $z$, one may write  
the wave functions in the adiabatic approximation \cite{Lesovik86}
\begin{equation}
\Psi(x,y,z) =\sum\limits_i\xi_i(x,y;S(z)) \psi_i(z)\, ,
                                                              \label{yy0}
\end{equation}  
where $\xi_i(x,y;S(z))$ is a transverse wave function of the $i$--th channel 
with an eigenenergy $\epsilon_i(S(z)) \equiv \epsilon_i(z)$. 
The longitudinal function, $\psi_i(z)$, satisfies a one dimensional 
Shr\"odinger equation
\begin{equation}
[-(2m)^{-1} \partial_z^2 + \epsilon_i(z) +V_b(z)] \psi_i(z) = 
\epsilon\,  \psi_i(z)\, ,
                                                              \label{yy1}
\end{equation}  
where $V_b(z)$ is the localized tunneling barrier potential. 
Thus 
the problem is reduced to a one dimensional effective tunneling problem 
described by Eq.~(\ref{yy1}). (See also Fig. \ref{fig2}.) If the constriction is 
narrow enough it closes  most of the channels, since 
in these channels $\epsilon_i(z) > \epsilon_F$ for small enough $z$. 
Hereafter we shall assume that this is the case.
Even in the  open channels there is a partial 
reflection of electrons due to 
the presence of the barrier, which is characterized by the barrier's
reflection coefficients, $r_i$.
For $z$ and $z'$ outside the barrier region, by employing  Eq.~(\ref{w2}) and 
expressing the Green's functions in the WKB approximation,
 one finds  for the conductance kernel (cf. with  Eq.~(\ref{w9}) )  
\begin{eqnarray} 
&&g(z,z',\omega)=  \frac{e^2}{\pi\hbar}
\sum\limits_{i}\left[
\exp\left\{
i \omega \left| \int\limits_{z'}^z{dz\over v_i(z)}\right| 
\right\}          \right.                          \label{yy2}    \\  
&&\left. - |r_{i}|^2 
\exp\left\{
i \omega \left| \int\limits_{0}^z{dz\over v_i(z)} \right|+
i \omega \left| \int\limits_{0}^{z'}{dz\over v_i(z)} \right| 
\right\}   \right]  \, , 
                                                              \nonumber
\end{eqnarray}
where 
\begin{equation}
v_i(z) = \sqrt{ \frac{ 2(\epsilon_F - \epsilon_i(z))}{m} } \, ,
                                                              \label{yy3}
\end{equation}  
and the summation runs over the open channels only. This expression will be 
used to find the four--terminal conductance of the barrier, measured by probing 
the voltage drop across it. (In very narrow energy intervals, 
when one of the channels is 
extremely near its transmission threshold it may not be tractable by WKB.)

Let us first imagine solving the problem on the large length scale, $|z|\gg L$  
\cite{foot6}, where the 
entire constriction may be treated as a single localized scatterer.   
The results of previous sections are directly applicable to such a problem 
(see e.g. Sec. \ref{s32}).
One finds that at sufficiently low frequencies the induced electric field, 
$E^{ind}$, is  confined to the constriction region. 
A line integral over this field 
is  a total voltage drop across the constriction 
\begin{equation}
\int\!\! dz E^{ind}(z,\omega) = V_{0}  \, ,
                                                              \label{yy4}
\end{equation}    
where the integral effectively runs over the region 
between the two points A and B (see Fig. \ref{fig2}a) located outside the 
constriction at a distance which 
is large compared with a bulk screening length (but still much smaller than $L$).   
Since only a few 
channels are open, the current is  determined by Eq.~(\ref{eB2}) and given by
\begin{equation}
I_0 \approx V_{0} \frac{e^2}{\pi\hbar}\sum\limits_i |t_i|^2   \, .
                                                              \label{yy5}
\end{equation}
There are corrections to this expression of the order of a ratio of number of open
channels to a total number of channels \cite{Landauer89}.     
Only the open channels contribute to the sum on the right hand side; the 
transmission coefficients of the open channels are those of the localized barrier.

To determine the voltage drop across the barrier itself one must 
determine  the structure of $E^{ind}(z,\omega)$ on a small length scale, 
which turns out to be the screening length, $\kappa^{-1}(\ll L)$. 
The induced field $E^{ind}$ is comprised by  the two components,
$E^{ind}=E^{ind}_1 + E^{ind}_2$, created by charge densities induced   
at spatially well separated locations. $E^{ind}_1$ is created by charges 
accumulated in the regions of the narrowing of the constriction. This field component 
is practically constant between points C an D in the constriction 
(see Fig. \ref{fig2}a).  Its contribution 
to the voltage drop measured across the barrier is small in the parameter 
$z_v/L \ll 1$, where $2z_v$ is a distance between the voltage probes. 
The other component, $E^{ind}_2$, is induced by charges 
accumulated near the localized barrier. It is sharply peaked 
in the barrier region and is primarily responsible for  
the voltage measured  across the barrier. 
The induced field, $E^{ind}(z,\omega)$,  inside the constriction region  
is given by a solution of the self--consistent equation (\ref{e2}), with the 
conductance kernel Eq.~(\ref{yy2}). 
For low frequencies, $(\omega L /v_F) \ll 1$, one may expand the exponents in 
Eq.~(\ref{yy2}) it terms of this small parameter and obtain to  leading order 
\begin{eqnarray} 
&& \frac{\hbar S(z)}{4 e^2}\, \partial_z E^{ind}(z) -
\left(\!\sum\limits_i{1\over v_i(z)}\! \right) \!\!
\int\!\! d z' \mbox{sign}(z-z') E^{ind}(z') 
           \nonumber  \\
&&= - \left(\sum\limits_i{|r_i|^2 \over v_i(z)} \right) 
\mbox{sign}(z) V_{0} \, ,     
                                                              \label{yy6}
\end{eqnarray}
where Eq.~(\ref{yy4}) was employed on the right hand side.
In the region where the effective potential is 
``flat'' (see Fig. \ref{fig2}d) 
and where $v_i(z) \approx v_i=\mbox{const}$ and 
$S(z) \approx S = \mbox{const}$, the solution of Eq.~(\ref{yy6}) is 
\begin{equation}
E^{ind}(z) = \mbox{const} +  
V_{0}\, \frac{\sum\limits_i |r_i|^2 v_i^{-1}} {\sum\limits_i v_i^{-1}}
\, \frac{\exp\{-|z|\kappa\} }{2\kappa^{-1}} \, ,
                                                              \label{yy7}
\end{equation}  
where the screening length is 
\begin{equation}
\kappa^{-1} = \sqrt{\frac{\hbar S}{8e^2\sum_i v_i^{-1} }} =
\frac{v_F}{\tilde \omega_p}\, . 
                                                              \label{yy8}
\end{equation} 
The constant on the right hand side  of Eq.~(\ref{yy7}) 
represents $E^{ind}_1$, the component created by the 
distant charges of the constriction's narrowing. Its value may be estimated
from Eqs.~(\ref{yy4}) and  (\ref{yy7}) as 
\begin{equation}
E^{ind}_1 \approx \mbox{const} \approx
\frac{V_{0}}{L} \, 
\frac{\sum\limits_i |t_i|^2 v_i^{-1}} {\sum\limits_i v_i^{-1}}\, . 
                                                              \label{yy7a}
\end{equation}

The voltage drop across the barrier, measured by the probes located 
just outside the barrier at points E and F (see Fig. \ref{fig2}a) with coordinates 
$z= \pm z_v$,
is given by $V=\int_{-z_v}^{z_v} dz E^{ind}$. Employing  Eq.~(\ref{yy7}),
one finds 
\begin{equation}
V=   
V_{0}\left[ 
\frac{\sum\limits_i |r_i|^2 v_i^{-1}} {\sum\limits_i v_i^{-1}}
+ O\left(\frac{z_v}{L},e^{-z_v\kappa} \right)\right]\, .
                                                              \label{yy9}
\end{equation}  
Therefore in the asymptotic regime $\kappa^{-1} \ll z_v\ll L$, one finds for the 
four--terminal conductance $\tilde g = I_0/V$,  
\begin{equation} 
\tilde g = \frac{e^2}{\pi\hbar} \, 
\frac{\sum\limits_i v_i^{-1} \sum\limits_i |t_i|^2}
{\sum\limits_i v_i^{-1} |r_i|^2}\, ,
                                                              \label{yy10}
\end{equation}  
in agreement with BILP \cite{Buttiker85}. 
We therefore confirm their result for the barrier embedded into the constriction 
attached to the wide reservoirs if the inequalities 
$\kappa^{-1}\ll L < v_i/\omega, l^{el}$ are satisfied.

Eq.~(\ref{yy10}) predicts a sudden drop of the conductance if one (or several) 
$v_{i_0} \to 0$. 
The physical reason for such behavior is an enhanced screening by the 
channel(s) with the small velocity and hence a large density of states. 
We stress that although a non--monotonic  
dependence of the conductance on the chemical potential may indeed take place, 
there is no actual discontinuity when  a channel opens.  
The reason is that the above calculations resulting in Eq.~(\ref{yy10}) 
lose their validity if $v_{i_0} < \hbar/(mL)$ \cite{foot2}  
(or $v_{i_0} < L\omega$). The two key 
approximations simultaneously go wrong in this case: (i) the long--wavelength limit 
for the non--local conductance, which neglects Friedel oscillations at $q=2k_F$ 
and (ii) the WKB approximation, which neglects reflection by $\epsilon_{i_0}(z)$, 
the  effective barrier due to the constriction. 
If these two effects are properly 
taken into account the continuity of conductance is restored, however 
its value for $0< v_{i_0} < \hbar/(mL)$ depends both on the 
position of the probes  and the specific shape of the   constriction.

Also, at the beginning of this Section, we assumed that if a channel is 
``closed'', i.e. $v_i \rightarrow iv_i$, the corresponding current vanishes. 
In fact this requires negligible tunneling or $|v_{i_0}| > \hbar/(mL)$.
Thus Eq.~(\ref{yy10}) also fails just below an energy threshold. 

Finally we address the frequency dependence of the four--terminal conductance. To 
this end we expand  Eq.~(\ref{yy2}) to the next order in $\omega L/v_F$ and 
substitute into the self-consistent equation (\ref{e2}). 
This gives the frequency dependent correction to the 
induced electric field  
\begin{equation}
\delta E^{ind}(\omega, z) = i\omega  
\frac{\sum\limits_i |r_i|^2 v_i^{-2}} {\sum\limits_i v_i^{-1}}
\, \frac{\exp\{-|z|\kappa\} }{2\kappa^{-1}}  
\int\!\! dz |z| E^{ind}(z)\, ,
                                                              \label{yy11}
\end{equation}  
where $E^{ind}(z)$ is the zero frequency field given by 
Eqs.~(\ref{yy7}), (\ref{yy7a}). According to Eqs.~(\ref{eA4}), (\ref{eA5}) 
the relation between the divergence-less current 
and the total voltage drop,  Eq.~(\ref{yy5}), should be modified to include 
the classical capacitance of the constriction, $C_0$.  
The resulting four--terminal conductance acquires a capacity--like 
frequency dependent correction of the following form  
$\delta \tilde g(\omega)  = - i\omega(C_0 + C)$, where  
\begin{equation}
C = \tilde g\,
\frac{\sum\limits_i |r_i|^2 v_i^{-2}} {\sum\limits_i |r_i|^2 v_i^{-1}}
\frac{\int\!\! dz |z| E^{ind}(z)} {\int\!\! dz E^{ind}(z)} \, .
                                                              \label{yy12}
\end{equation} 
The precise value of the last factor in this expression depends on 
$E^{ind}(z)$ on the scale $z\sim L$ and hence on the shape of the constriction. 
It may be estimated using Eqs.~(\ref{yy7}), (\ref{yy7a}). 
Provided that  
$(L\kappa)^{-1} < |t_i|^2$, which is valid for all barriers except those 
with almost complete reflection, one finds for the additional capacitance  
\begin{equation}
C= \gamma L\, \frac{e^2}{4\pi\hbar}\sum\limits_i |t_i|^2 \,
\frac{\sum\limits_i |r_i|^2 v_i^{-2}\sum\limits_i |t_i|^2 v_i^{-1}} 
{\left[\sum\limits_i |r_i|^2 v_i^{-1}\right]^2}\, , 
                                                              \label{yy13}
\end{equation}
where $\gamma = O(1)$. 
In the opposite limit $|t_i|^2 \ll (L\kappa)^{-1}$ one obtains 
\begin{equation}
C = \kappa^{-1}\, \frac{e^2}{\pi\hbar}\sum\limits_i |t_i|^2 \,
\frac{\sum\limits_i |r_i|^2 v_i^{-2}\sum\limits_i v_i^{-1}} 
{\left[\sum\limits_i |r_i|^2 v_i^{-1}\right]^2}\, . 
                                                              \label{yy14}
\end{equation}
We stress the peculiar dependence of the additional capacitance 
on channel velocities and in particular a sharp {\em rise} at the point of a new  
channel opening. 
(As was mentioned above, there is no actual discontinuity, since 
Eqs.~(\ref{yy13}), (\ref{yy14})
are not applicable once $v_{i_0} < \hbar/(mL)$.)

\section{Discussion of the results}
\label{s4}

We have presented an approach to the Landauer four--terminal 
tunneling conductance, which does not assume the presence of two reservoirs which  
maintain different chemical potentials. Our theory is based on the solution of a 
self--consistent set of equations involving the non--local conductance, 
$g(z,z',\omega)$, the tunneling current, induced 
charge density and electric field. 

\begin{figure}
\vglue 0.5cm
\hspace{0.01\hsize}
\epsfxsize=0.9\hsize
\epsffile{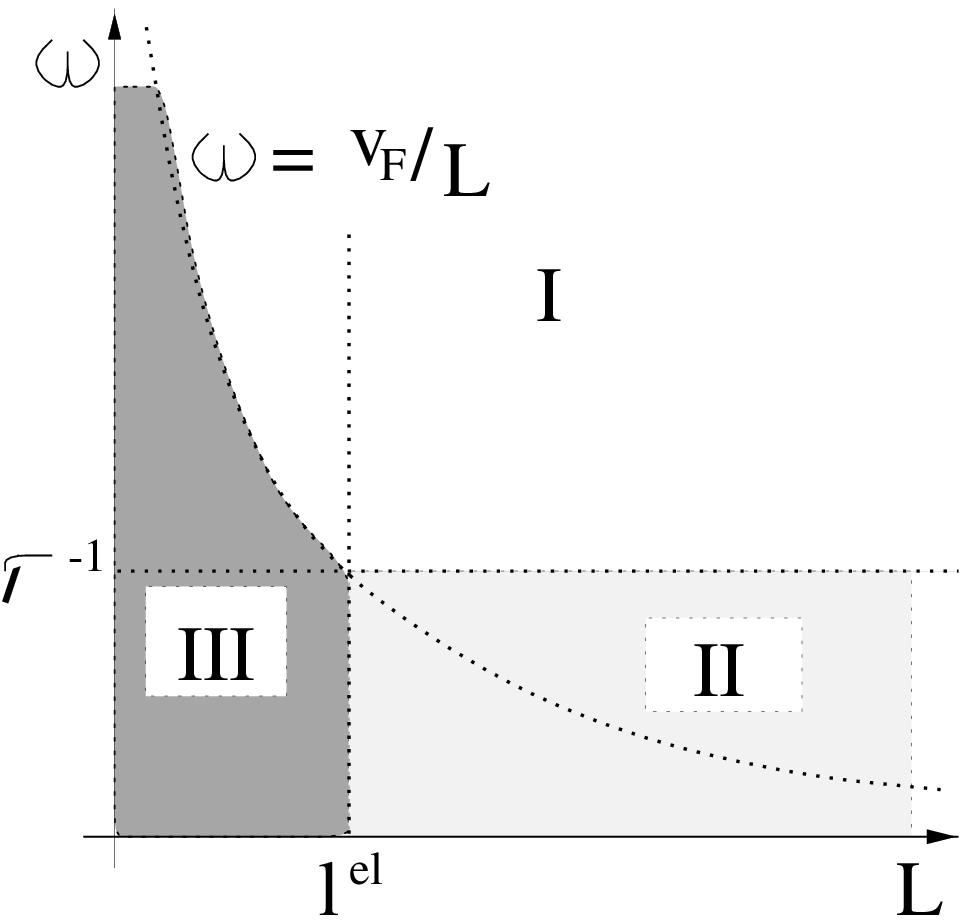}
\refstepcounter{figure} \label{fig4}
\vspace{0.1\hsize}
{\small FIG.\ \ref{fig4} Various physical regimes; 
area I corresponds to a  ``high'' frequency regime, 
II to a ``disordered leads'' and  
III to a ``ballistic constriction''  geometry.
Here $L$ is the length of the leads attached to the barrier, $\omega$ -- the 
frequency of the driving field, $l^{el}$ and $\tau^{-1}$ are elastic mean free
path and scattering rate correspondingly.  
 \par}
\end{figure}

There are three different regimes (see Fig. \ref{fig4}), 
defined by the length of the leads,
$L$,  the frequency, $\omega$, and the rate of elastic scattering, $\tau^{-1}$
(or the elastic mean free path, $l^{el}= v_F \tau$). 
The first ``high'' frequency regime denoted by I in Fig. \ref{fig4} 
is defined by the conditions 
$\omega > \mbox{max}\{\tau^{-1}, v_F/L\}$. We have treated it in  section \ref{s2}.
The second denoted by II is that of ``disordered'' leads. It is defined by 
the conditions $L > l^{el};\, \omega < \tau^{-1}$ and was treated in 
section \ref{s5}. Finally the third one denoted by III and defined by 
$L < l^{el};\, \omega < v_F/L$ is the regime of 
``ballistic'' motion. It was considered in  section \ref{s7}.

A few remarks about the relationship of our work in the ballistic regime with a 
barrier, in the geometry of Fig. \ref{fig2}a, and the work of BILP 
\cite{Buttiker85}: First, the final result (our Eq.~(\ref{yy10}) ) is the same in 
both treatments. However: in BILP the occupation of electronic states in the 
presence of a current is different in the regions to the left and to the right 
of the barrier. In our work the presence of a current does not modify the 
occupation of the electronic states (which extend from the left through the barrier 
to the right), but does modify their wave-functions -- in accordance with standard 
Kubo transport theory.  BILP's chemical potential differences $\mu_1-\mu_2$
and $\mu_A-\mu_B$ are respectively equal to the electrostatic potential 
differences $V_0$ and $V$ of Fig. \ref{fig2}c. We have not been able to derive 
the very simple picture of BILP from our considerations. 
 
In the regimes I and II we have obtained  results which are qualitatively 
different from ballistic regime III. The most striking 
difference occurs for the weakly reflecting  
barrier (cf. Eqs.~(\ref{q4}) and (\ref{q3})). In particular, 
for  long and/or disordered leads the conductance does not 
exhibit discontinuities when the velocity of one of the 
channels vanishes.    The physical reason for these differences lies in the 
difference of  momentum distribution functions in the diffusive and high frequency 
regimes compared to that in the ballistic regime.  
In the opposite limit of  weak transmission one finds, to  leading order, 
the same result for all three regimes, which is identical with  
Landauer's two--terminal expression, Eq.~(\ref{q1}).   
We believe that the predicted qualitatively different dependence 
on the channel velocities (and hence on the chemical potential) 
for  diffusive and ballistic leads may be checked experimentally.

We also predict a peculiar dependence of the capacitance on the 
chemical potential, Eqs.~(\ref{yy13}), (\ref{yy14}), 
which, we hope, may  also be checked experimentally.

\section{Acknowledgments}
\label{s6}

We are indebted to Rolf Landauer and Joe Imry for illuminating discussions. 
This research was supported by the NSF grant  DMR 96--30452. 
A.K. gratefully acknowledges the hospitality of the Institute for 
Theoretical Physics during his stay in Santa Barbara.

\widetext 

\begin{thebibliography}{10}
\bibitem{foot1} One has to be aware
that the full expression for $g(z,z',\omega)$ contains also $G^RG^R$ and 
$G^AG^A$ terms along with the diamagnetic contribution. These terms, however, 
cancel each other exactly in the leading order 
in $\omega/\epsilon_F$. Since only the 
low frequency conductance is the subject of the present paper, we do not 
discuss these terms here.
\bibitem{foot2} To shorten the subsequent expressions we do not distinguish 
between the left and the right reflection coefficients, which corresponds to a
symmetric barrier. All the finite results, however, are valid for an asymmetric 
barrier as well.
\bibitem{foot6} It is assumed hereafter that the constriction is characterized by 
a single length scale, $L$. Generalization for a constriction described by several 
length scales is straightforward.
\end{thebibliography}
\end{document}